\documentclass[fleqn,twoside]{article}
\usepackage[headings]{espcrc2}

\readRCS
$Id: espcrc2.tex,v 1.2 2004/02/24 11:22:11 spepping Exp $
\usepackage{graphicx}
\usepackage{epsfig}

\usepackage[figuresright]{rotating}

\newcommand{\AmS}{{\protect\the\textfont2
  A\kern-.1667em\lower.5ex\hbox{M}\kern-.125emS}}

\title{The origin of the ankle}

\author{Antonio Codino\address{ INFN and Dipartimento di Fisica dell'Universita
        degli Studi di Perugia, Italy } 
        and
        Francois Plouin\address{ Former CNRS researcher, Ecole Polytechnique, LLR, 
        F-91128 Palaiseau, France}}

\begin{document}

\begin{abstract}
({\it {\b  to appear in the CRIS 2006 proceedings}})

The differential intensity of cosmic radiation shows a sequence of
depressions  referred to as
$knees$ in a large energy band above $10^{15}$ $eV$. The global
depression entailed in the complete spectrum with
respect to the extrapolated intensity based on low energy data amounts to a
maximum factor of 8, occurring at $5 \times 10^{18}$ $eV$, where flux measurements
exhibit
a relative minimum, referred to as the $ankle$. It is demonstrated by a full
simulation of cosmic ray trajectories in the Galaxy that the intensity minimum 
 around the ankle energy is primarily due to the nuclear interactions
of the cosmic ions with the interstellar matter
and to the galactic magnetic field. $Ankles$ signal the onset energies  of
the rectilinear propagation in the Milky Way at the Earth, being for example,  
$4 \times 10^{18}$ $eV$  for helium and $6 \times 10^{19}$ $eV$  for iron.
The ankle, in spite of its notable importance at the Earth, is a local 
perturbation
of the universal spectrum which, between the knee and the ankle,
decreases by a round factor $10^{9}$,  regaining its unperturbed status
above $10^{19}$ $eV$.
\end{abstract}

\maketitle

\section{INTRODUCTION}

Energy measurements of giant air showers demonstrated the existence of
a distinctive structure, the $ankle$, in the differential energy spectrum of the 
cosmic radiation above $10^{18} eV$. Figure 1 shows the energy spectrum measured
by four experiments [1,2,3,4]
at these extreme energies. 
Cosmic ray intensity  does not continue to decrease
with a spectral index between 3 and 3.2, observed between $10^{16}$ and 
$10^{18} eV$, but an enhancement of the intensity above $5 \times 10^{18}$ $eV$
appears in all experiments.
This enhancement is relative to the  extrapolation at high energy with
the index  of 3 measured at lower  energy, below $10^{18}$ $eV$.
The $ankle$ is the distinctive pattern in the spectrum 
consisting of a minimum of intensity followed by an enhancement observable 
approximately in the energy band $5 \times 10^{18}$-$7 \times 10^{19} eV$.  
The energy at which this enhancement occurs, 
 and its magnitude,  of this enhancement do not coincide
in different experiments, though precise measurements of the minimum are available from 
the Hires Collaboration [4,5]. 

This contribution to the CRIS Conference 2006 follows recent studies 
[6,7] (hereafter Paper 
I and II, respectively) reporting a   
quantitative  solution of the long-standing 
problem of the knee and ankle of the cosmic ray spectrum.
A notable aspect of this solution is that the same mechanisms are responsable
for  the ankles and the knees 
of the individual ions. Even purposely
it would be impossible to separate the explanation of the ankle from that
of the knee. 

The intrinsic mechanisms 
generating the ankles do not depend on the spectral indices of the cosmic rays 
at the sources nor on any extragalactic component of the cosmic radiation, but  is 
an intrinsic property of the galactic cosmic rays (Sections 3 and 4). The 
extragalactic component, if any at the Earth around $10^{19}$ $eV$, might add to the 
galactic component, affecting the magnitude of the ankle,
but its necessity and its experimental evidence remain unproved (Section 7). Note that 
the expected GZK depression 
or its absence [3] is
beyond the energy band pertinent for  the explanation of the ankles
presented here.

\section{COSMIC ION TRAJECTORIES}

\begin{figure}[h]     
\epsfig{file= 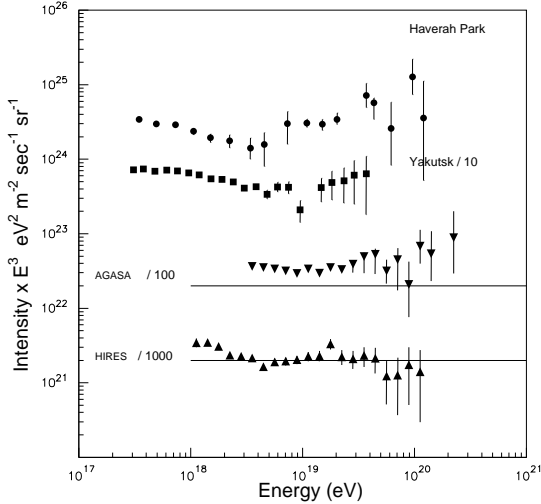,width=85mm}
\caption{Cosmic ray spectra in the energy band around $10^{17}$- 
$10^{20}$ $eV$ from Haverah Park, Yakutsk, Agasa and Hires experiments.
The two horizontal thin lines mark the discrepancy in intensity between the
Agasa and Hires experiments.}
\label{fig:fig1}
\end{figure}
The method of calculation, 
used for over a decade, reconstructs  ion trajectories by computer
simulations and evaluates physical quantities of cosmic rays (intensities,
energy  spectra, ion ratios etc) using 
the number of trajectories intercepting a given small volume in the Galaxy
which in the present calculation can be the Earth (also
$local$ $galactic$ $zone$ or $solar$ $cavity$), a sphere concentric to the Earth, or
the entire disc volume (Section 9). 
The  computational algorithms, described elsewhere [8,9,10] take into account the 
following astronomical, astrophysical and radioastronomy observations:
(1) the spiral magnetic field;
(2) the field strength (see figure 3 Paper I) of the spiral magnetic field;
(3) a chaotic magnetic field with a field strength of about three times that
of the regular field;
(4) the form and the dimension of the Galaxy (see
    figure 1 in ref.[9]); 
(5) a uniform distribution of cosmic ray sources in the
    galactic disk (see eqn.(3) in ref.[9]);  
(6) the nuclear cross sections ion-hydrogen, $\sigma$;
(7) the interstellar matter density  in the disk, $d$,  of 1.24 hydrogen 
atoms per $cm^3$;   
 (8) the position of the $solar$ $cavity$ inside the 
disk, at 14 $pc$ above the galactic midplane and 8.5 $kpc$
from the galactic center; (9) the galactic wind (Section 2 in ref.[10]). 
Prerequisites of this study regarding the notion  of $galactic$ $basin$ [11,10]
have been previously discussed in detail. 

A cosmic ion emanated from a source travels, on average,  a mean length, $L$, 
in the disk volume, primarily determined by $\sigma$
and $d$. Typically, this global
length $L$ is subdivided into thousands and thousands of segments, depending on
the particular region of propagation and the energy of the cosmic ion. The ion 
propagation normal to the regular spiral field is generated by the chaotic
field leading to a transverse-to-longitudinal displacement ratio 
compatible with the quasi-linear theory of ion propagation in an
astrophysical environment.
The chaotic or turbulent magnetic field
is materialized by magnetic cloudlets
(see, for instance, ref.[12,13])  with variable dimensions
and a field strenght three times
greater than that of the regular field.

Trajectories at low energy are quite different 
from those at very high energy. Figures 2 and 3 display a typical low energy
trajectory.
\begin{figure}[h]     
\epsfig{file= 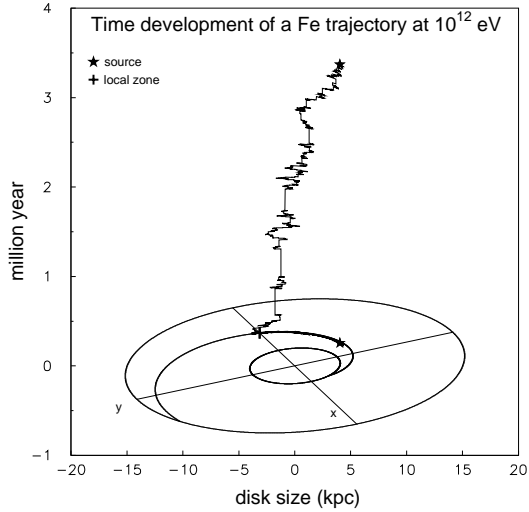,width=85mm}
\caption{Example of an iron trajectory in the disc and its time evolution
(vertical axis) during $3.5 \times 10^{6}$ $years$.
Note that the zig-zag, caused by the turbulent field, disappears 
for trajectories around and above $10^{18}$ $eV$, the rectilinear
propagation region.}
\label{fig:fig2}
\end{figure}
\begin{figure}[h]    
\epsfig{file=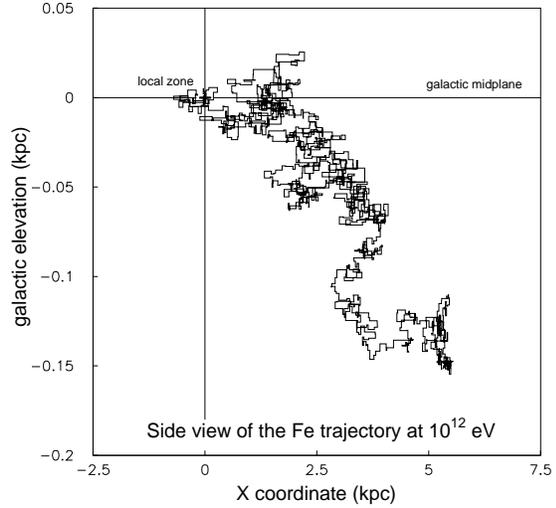,
width=85mm}
\caption{ Side view of the 
same iron trajectory  shown in figure 2, with a length of $334$ $098$ $pc$
and a straight-line distance between source and Earth of 671 $pc$. 
The star denotes the Fe source at $x$=$5.404$, $y$=$2.826$ and $z$=$-0.148$
$kpc$ while
the cross shows the Earth's position.}
\label{fig:fig3}
\end{figure}
\begin{figure}[h]  
\epsfig{file=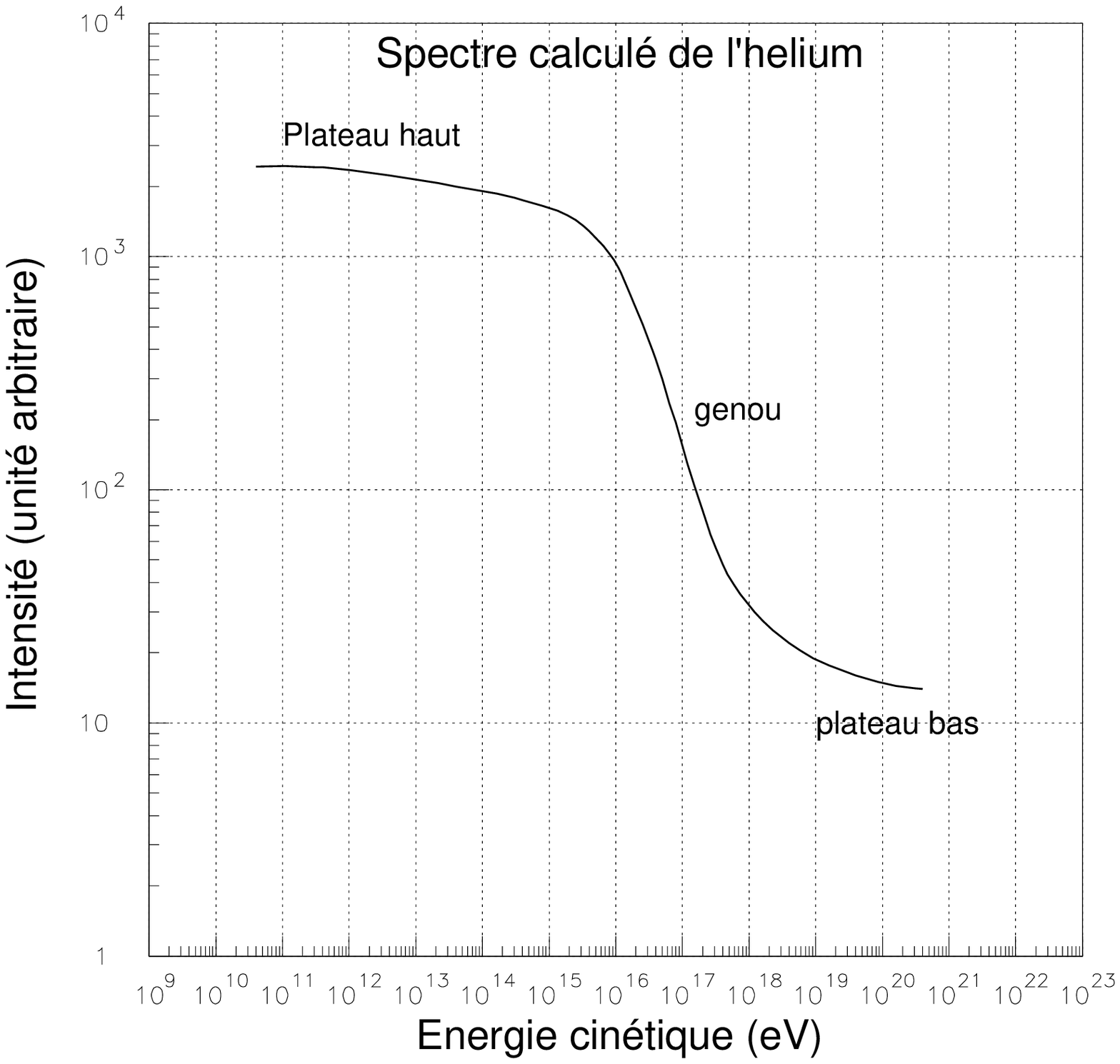, width=85mm} 
\caption{ Number of cosmic rays (He) reaching the local galactic zone, $n_g$,
versus energy. This quantity is related to helium intensity via 
the spectral index of cosmic helium at Earth.
The structure of the curve exhibits a high plateau, a rapid descent and a low 
plateau.} 
\label{fig:fig4}
\end{figure}

\section{WHAT ARE THE KNEES OF THE INDIVIDUAL IONS}

The differential energy spectrum of the cosmic rays,  $dn$/$dE$, depends on 
the number of cosmic ray trajectories intercepting the local
galactic zone, $n_g$. Figure 4 shows $n_g$ versus energy for helium,
taken as an example.
Three energy regions of $n_g$ are clearly distinguishable: (1) a high plateau; 
(2) a rapid descent; (3) a
low plateau.  
The intensity gap  between the high and the low plateau is of capital
importance to comprehend the origin of the ankle. This gap depends on the
matter thickness encountered by the cosmic helium in the Galaxy 
and on the  nuclear cross section helium-hydrogen, $\sigma$(He).
\begin{figure}[h]     
\epsfig{file=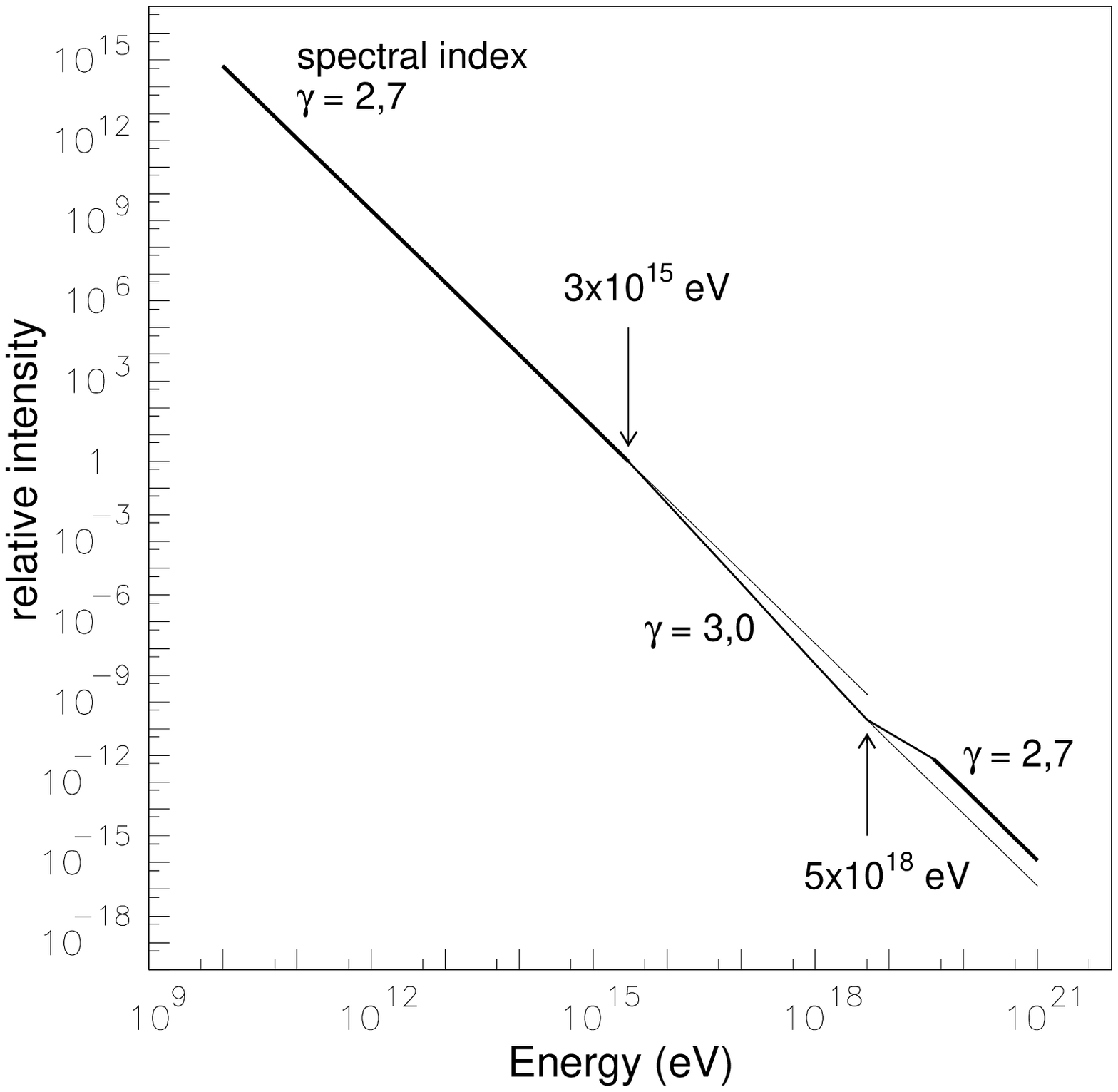,width=85mm}
\caption{Cosmic ray spectrum  measured at Earth in the interval $10^{10}$-  
$10^{20}$ $eV$ with the relevant spectral indices of 3 and 2.7 [3] after the 
knee and ankle energies, respectively. 
The approximate equality of the spectral index over the enormous
energy range suggests that a universal acceleration 
engine operates inside and outside galaxies 
as argued elsewhere [14].}
\label{fig:fig5}
\end{figure}
\begin{figure}[h]    
\epsfig{file=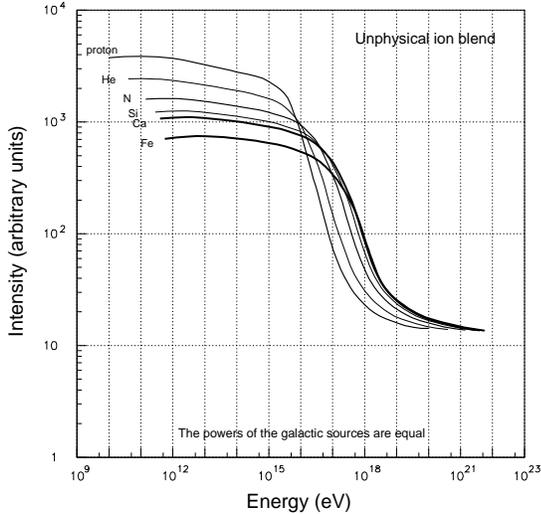,width=85mm} 
\caption{Energy spectra of six nuclear species obtained by ideally
assuming equal source powers of any galactic ions called here
unphysical ion blend. This ideal spectra transform into real, physical
spectra, shown in figure 7, once source abundances and spectral indices
are considered.} 
\label{fig:fig6}
\end{figure}
\begin{figure}[h]    
\epsfig{file=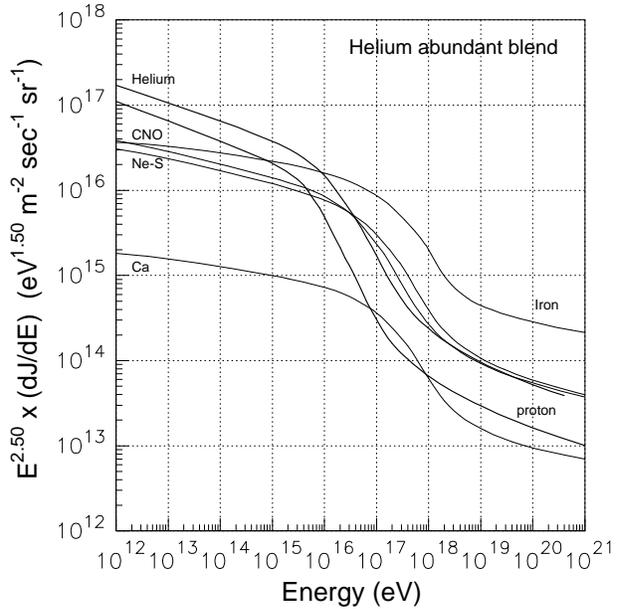,width=85mm}
\caption{Computed energy spectra of individual ions according
to the elemental abundances and spectral indices of the blend 3 of table 1. 
These curves are just one of the many examples of how the energy spectra 
given in figure 6 may be transformed by a particular ion blend.} 
\label{fig:fig7}
\end{figure}
\begin{figure}[h]    
\epsfig{file=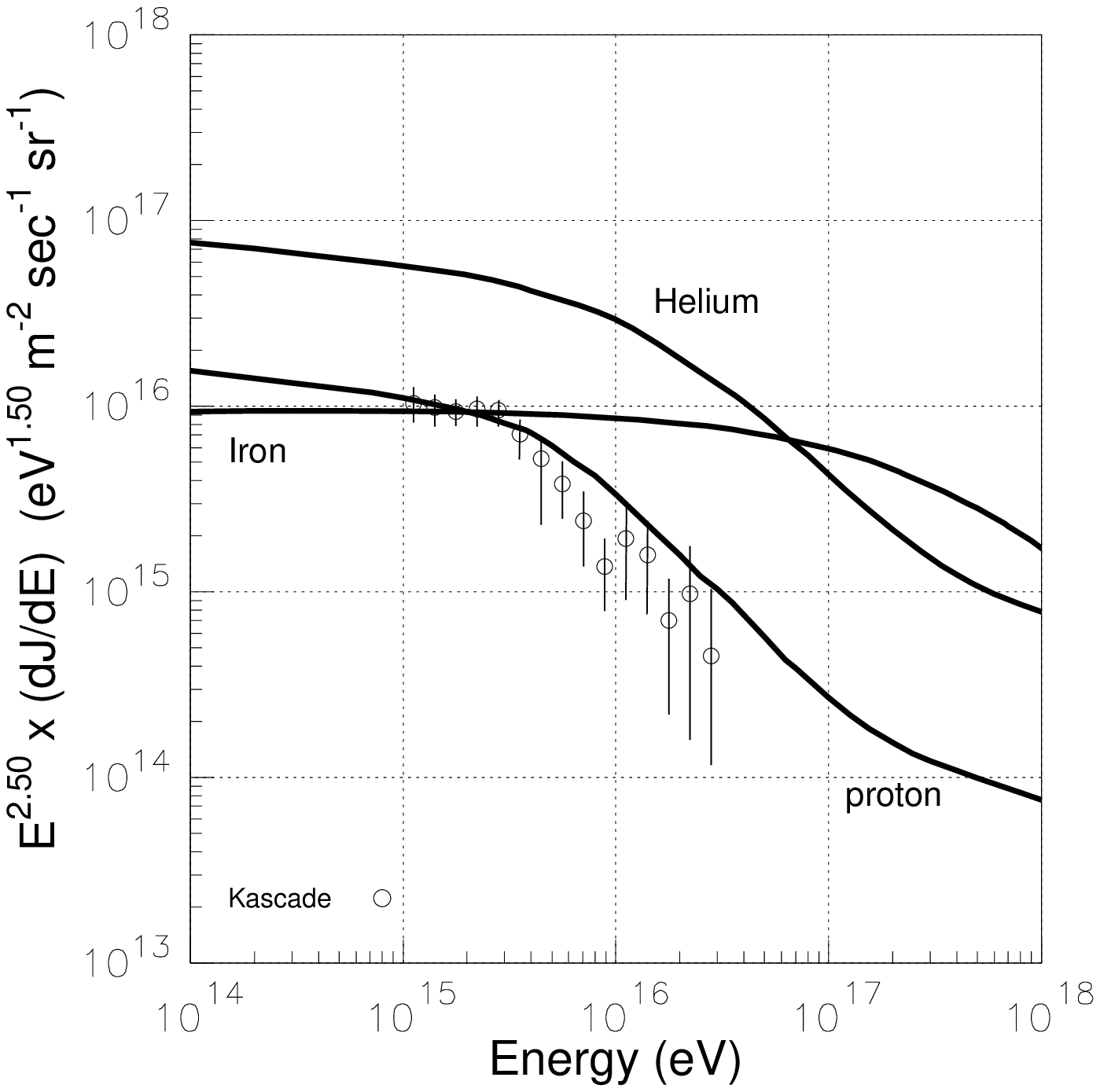 ,width=85mm}
\caption{Comparison between computed and measured proton spectrum.
The computed He and Fe spectra with  
indices of 2.6 (He) and 2.5 (Fe) are  also shown as a relevant grid of ion 
fluxes. The spectra normalization assumes flux equality for proton and iron at 
$10^{15.3}$ $eV$ and a  
$He$/$Fe$ flux ratio different from that of table 1.} 
\label{fig:fig8}
\end{figure}

The form of the spectrum shown in figure 4 is quite general, 
and simple mechanisms concur in its creation as clarified below. 

In the energy band $10^{10}$-$10^{15}$ $n_g$ is approximately
constant, though slightly decreasing. Forcing a power law interpolation
on $n_g$ with a constant spectral index,  a value of 
0.13 for helium results. This decreasing feature is due to rising cross sections
as explained in detail in Paper I, Sections 3 and 4. In the high
plateau region,  the indices of $n_g$ for different nuclides
tend to decrease with the atomic number. An elegant derivation of this result 
is feasible by the notion of $galactic$ $basin$ [10,11]. 

The second portion of the spectrum,  referred to as the knee,
is due primarily to the break in the efficiency of the ion bending inherent the 
galactic magnetic field. This break has been analyzed in detail elsewhere 
(Paper I Section 4 and Paper II Section 5). 
The break energy is defined as that particular energy characteristic  of a nuclear
 species, where the bend of $n_g$ occurs and a sharp decrease dominates. 
Though the effect of the 
magnetic field on $n_g$ is dominant, the exact form of $n_g$ emerges only
after including the position of the solar cavity,
the nuclear cross sections and the disc size.
The decrease of the cosmic ray intensity,  or equivalently
the descent in the bending efficiency and comcomittant effects on $n_g$
of the other parameters, cannot continue
indefinitely as the energy increases above the break energies. 
In fact, the average value of the
magnetic field strength is finite (say, 3 $\mu$G), and because of this
finiteness, at some energy,
for a specified ion, trajectories remain completely unbent, and consequently, the bending 
efficiency vanishes. This condition corresponds
to the rectilinear propagation which is at the origin of the $ankles$
as discussed in Section 8. 

\section{POSTULATING CONSTANT SPECTRAL INDICES IN THE COSMIC RAY SPECTRUM}

The existence of the ankle is incubated in 
the quantity $n_g$ versus energy shown in figure 4.  Subsequent 
analysis indicates how the low plateau in figure 4  
transforms into a physical $ankle$ accessible to experimentation. 
Figure 5 shows an idealization of the universal spectrum of the 
cosmic radiation between  $10^{10}$ and $10^{21}$ $eV$.
A notable feature of this spectrum
is its linearity in logarithmic scales. 
The ankle is a small perturbation of the differential intensity. 
In fact the maximum difference between extrapolated
and measured intensity is  a factor 8 occurring around $5 \times 10^{18} eV$ 
(see figure 20 in Paper I and the data in ref.[4]), while in the knee-ankle energy
range the differential intensity falls by a round factor $10^{9}$, hence
the term $small$ $perturbation$. A similar conclusion is reached in ref.[15]  
where other empirical arguments are used.  The remarkable linearity shown in 
figure 5 suggests that
individual ions may also have approximate linearities. 
For logical coherence in the explanation of the ankle this linearity 
is converted here in a postulate.
Notice that this postulate constitutes a useful $ordo$ $rerum$ in the experimental
data below $10^{15} eV$ and above $10^{17} eV$ for a number of reasons [16].
Also, the quantitative account of the knee and the ankle given here
would demand a much softer hypothesis than this powerful postulate. For instance, 
only spectral indices in narrow energy bands, before and close the knee, are really 
useful in the present analysis. 
The absence of this postulate does entail other silent hypotheses (usually 
unexpressed in the literature). The rejection of the postulate
would entail more complex hypotheses, probably with higher arbitrariness.
The experimental data at low energy
indicate that spectral indices occasionally are not constant,
as reported, for example,  by the Atic experiment [17] for a number of ions 
between $10^{12}$ and  $10^{13} eV$ 
(C, Mg, O, Si and probably K) but a 
clear spectral break is not experimentally observed below the knee region 
as pointed out by others [18].   

The alteration of 
$n_g$ caused by the spectral indices does not spoil the existence of the ankle
already present in $n_g$ versus energy as a low plateau (see figure 6). Nor 
could the particular
values of the spectral indices of all ions influence the existence the ankle 
since the sum of the individual ion spectra, on average, should compensate, 
in order to recover imperatively ,the linearity shown in figure 5.  
How the function $n_g$ is altered by the particular spectral indices 
is illustrated in figure 7 for helium with an index of 2.72, and also for
other ions.

\section{COMPUTED AND MEASURED INDIVIDUAL KNEES}

The indices and elemental abundances used in the present calculation
are  shown in Table 1,  inspired from the results of some experiments
(Kascade, Jacee, Atic, Runjob, Eas-top and others).
Figure 8 shows the proton energy spectrum with a spectral index of 2.6,
based on the Atic experiment [17]
, at very low energy. This spectrum is normalized 
to the flux  of  $0.97 \times 10^{-22}$ $particles$/$eV$ $m^2$$sr$ $sec$,  
measured by Kascade (Sibyll) [19,20]
at the energy of $ 2 \times 10^{15}$ $eV$.
The computed spectrum agrees fairly well with the measurements; 
taking a softer index (for example, 2.72 instead of 2.6)     
the agreement with the experimental data becomes excellent. The energy
spectrum  obtained by the QJSjet algorithms in Kascade gives a similar accord,
since the
resulting spectrum rigidly shifts in intensity. Note that
this shift
in intensity is absorbed in the normalization and the accord persists.
The computed proton spectrum seems in disagreement with that measured 
by the TibetAS$\gamma$ Collaboration [21]
which has a lower energy bend and a softer
descent compared to these calculations and to the Kascade data. 

The computed helium and iron spectra also 
exhibit a good
agreement with the Kascade and Eas-top data 
as discussed in detail in Paper I and II.

Let us remark here that by altering the values of
the ion blends, within plausible empirical limits,
the accord between computed and measured spectra
persists.
\begin{table}[t!]
\begin{center}
\caption{ Relative abundances of cosmic ions and spectral indices
at the arbitrary energy of $2 \times 10^{15}$ $eV$ adopted in the
calculation.
The ion groupings CNO, Ne-S and Fe(17-26) derive from the 
traditional data analysis in many experiments. The composition of 
the blend 3, not given in 
the table, is 19.6, 35.5, 14.0, 12.0, 17.8 and 0.1 with the same spectral
indices of the blend 2 except for the proton index which is 2.72.} 
\vspace{1pc}
\begin{tabular}{lllllll}
\hline
           & Blend 1  &         & Blend 2  &        \\
           &          &         &          &        \\
\hline
           & Proton   &         & Proton   &        \\
           & abundant &         & superabun&        \\
\hline
Ion        & Comp.    & Index   & Comp.    & Index  \\
           & p. cent  &         & p. cent  &        \\
\hline
H          & 32.8     & 2.72    & 37.9     & 2.74   \\
He         & 29.7     & 2.72    & 27.4     & 2.72   \\
CNO        & 11.7     & 2.65    & 10.8     & 2.65   \\
Ne-S       & 10.0     & 2.65    & 9.3      & 2.65   \\
Fe(17-26)  & 15.0     & 2.60    & 13.8     & 2.50   \\
Ca         &  0.8     & 2.60    &  0.8     & 2.60   \\
\hline
$\gamma$   &          & 3.05    &          & 3.06   \\
           &          &         &          &        \\
\hline
\end{tabular}
\end{center}
\end{table}

\section{THE COMPLETE SPECTRUM BETWEEN THE KNEE
AND THE ANKLE}

The $complete$ $spectrum$ of the cosmic radiation is the sum of the 
$partial$ $spectra$ of the individual ions. Of course, it is assumed that
electrons, positrons, antiprotons and other elementary particles have negligible
fluxes compared to the global nuclide flux.

In figure 9,  
the computed energy spectra for the ion blend 2, in the interval 
$10^{10}$-$10^{21}$ $eV$,
are given. The Tibet [22]
 and Haverah Park data [1]
are reported in figure 9 while those of the 
Akeno and Agasa experiments [3] were shown previously (figure 12 Paper II).
The computed spectrum is required to agree with the experimental data
only at arbitrary energy point of $ 1.0  \times 10^{16}$ $eV$ 
where  the intensity ratio 
between Tibet and Kascade experiments is taken 1.4. 
Since no other constraint is imposed 
the overall accord is fairly good.
The spectral index of 3.06 for this ion blend 2 
in the energy interval $ 6.0  \times 10^{15}$-$10^{17}$ $eV$ is 
in excellent agreement
with the data of the quoted experiments and others not shown,
to avoid confusion in data superposition. The
ion blends 1 and 3 give  3.05 and 3.0, respectively,
for the global spectral index $\gamma$.
\begin{figure}[h]    
\epsfig{file=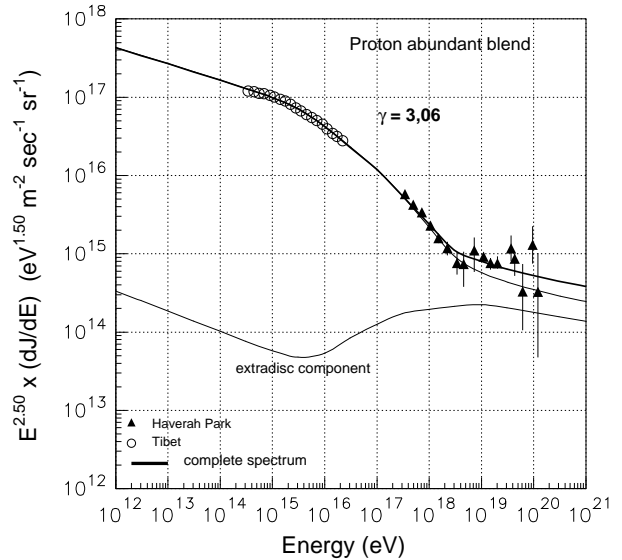,
width=85mm} 
\caption{Comparison between computed and measured spectra for the 
blend 2 given in table 1 and the data from the Tibet
and Haverak Park experiments covering the energy range $10^{14}$- 
$10^{20}$ $eV$.} 
\label{fig:fig9}
\end{figure}

\section{GALACTIC AND EXTRAGALACTIC COMPONENTS AROUND AND ABOVE THE
ANKLE ENERGY}

The extragalactic component of cosmic rays at the Earth, if it exists, should
cluster between $3.0 \times 10^{18}$ (proton ankle) and $5.0 \times 10^{19}$ $eV$
(iron ankle). 
This fundamental result, derived elsewhere (Paper II, Section 7), is almost 
independent of the method of calculation. Taking advantage of this outcome two
components of the cosmic ray flux at the Earth may be disentangled: galactic and 
extragalactic. The distinction in two components rests on the initial 
sites (Milky Way or outside) of the acceleration regions 
once secondaries (secondary protons, $^{3}$He, Be, B, etc.) generated in the wakes 
of the primaries may be neglected. 
The term $extragalactic$ would denote a number of possibilities:
(a) debris of normal galaxies in the cosmic vicinity; (b) reentrant particles
overflowing from the Milky Way; (c) debris of peculiar powerful galaxies;
(d) extragalactic particles of very high energy 
accelerated in space between galaxies. 
For the case (b) the more appropriate term $extradisc$ 
cosmic rays is used. 

The ratio between the galactic and extragalactic intensities at Earth, $I_g$/$I_e$,
in the band $3.0 \times 10^{18}$ to $5.0 \times 10^{19}$ $eV$ depends on the
ion blend adopted for the galactic component. Detailed calculations with 
the blend 2 (Paper II Section 6) give the energy spectrum of the extradisc
component, peaking around $10^{19}$ $eV$, as shown in figure 9. 
At this particular  energy, the  difference between the 
measured intensity and the computed galactic intensity is regarded, by definition,  
as an extradisc component. For the data shown in figure 9
the ratio $I_g$/$I_e$ is $2.7$ while the Agasa data with the same blend 2
give  $I_g$/$I_e$=$1.6$. 
 
\section{THE RECTILINEAR PROPAGATION AND THE ION ANKLES}

The galactic magnetic field is the basic element (though insufficient) for the 
comprehension of the processes generating the ankle. The form  
and strength of the
galactic magnetic field with its intrinsic turbulence are known with relatively
high precision, adequate for this study. The adequacy is justified elsewhere 
(Paper II Appendix B). The average  magnetic field strength 
fixes two milestones along the energy axis related to the alteration
of the ion trajectories in the Milky Way: the ankles occur at the energies
where the 
rectilinear propagation of cosmic ions commences,  while the knees are placed
at energies  where
the ion bending efficiency departs from an approximately uniform distribution, 
typical at lower energies.
By a uniform bending is meant an approximately isotropic distribution in the 
ion directions
in the vicinity of the source. The average distances travelled by cosmic 
ions in the Milky Way (source vicinity) have been previously calculated [10]. 

One should notice that the quantity
$n_g$ versus energy changes by more than two orders of magnitude from the low 
to the high plateau (see figure 4). This computed gap is close to physical 
reality because of the accord with the data displayed in figures 8 and 9.
The upper limits of the measured anisotropies in the arrival direction of 
cosmic rays  at the Earth around the knee energy are
about $10^{-3}$,  and recent measurements [23,24] indicate the persistence
of this figure even above $10^{18}$ $eV$ [5]. Consequently, the 
intrinsic mechanisms producing the knee dominate by several orders 
of magnitude those producing anisotropies. It follows that 
the accuracy required for knee and ankle calculations is, by far, less 
demanding that that required for anisotropy calculations. 
\begin{figure}[h]    
\epsfig{file=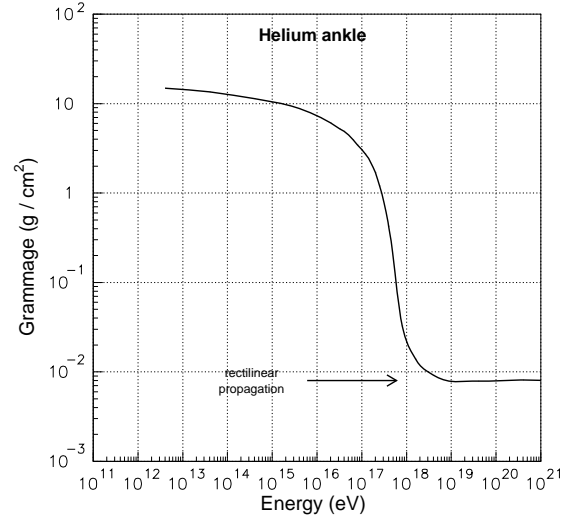,width=85mm}
\caption{ Helium grammage versus energy
in the galactic disc.
The arrow indicates the particular minimum energy of $ 4 \times 10^{18}$  $eV$ 
at which the grammage attains its asymptotic value. The shape of the grammage
agrees with that inferred ( via residence time) from measurements
of cosmic ray anisotropy versus energy [25] and it disagrees with the
unphysical, erroneous grammage extrapolated from $B$/$C$ flux ratio and other 
secondary-to-primary ratios 
measured at very low energy.} 
\label{fig:fig10}
\end{figure}
\begin{figure}[h]     
\epsfig{file=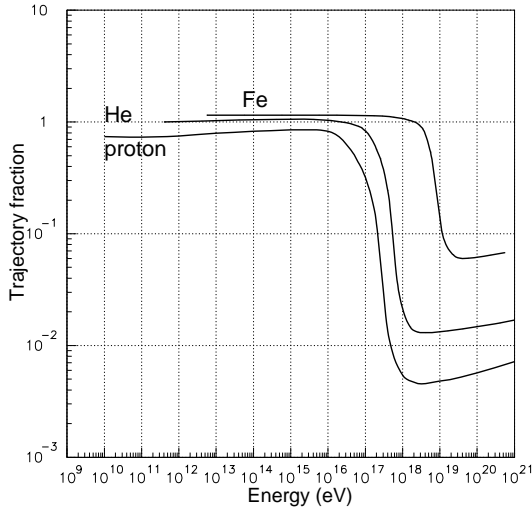,width=85mm}
\caption{Fractions of proton, He and Fe trajectories, $f_N$, emitted by ideal sources 
placed at Earth and terminated in the galactic disc by nuclear collisions. 
The He fraction  is normalized  to 1 at the energy of $10^{12}$ $eV$
and the Fe and proton fractions follow this normalization. These functions are 
related to the 
intensity of the extragalactic cosmic
rays intercepting the Earth by the arguments described in Section 4 of 
ref.[10].}     
\label{fig:fig11}
\end{figure}

For a given magnetic field configuration in the Galaxy,  the rectilinear
propagation of an ion occurs at a distinctive, particular energy.  Ion bending expressed by 
the curvature radius $R$ in a small volume permeated by a uniform constant field  
obeys the equation $R$=$p$/$Q$$B$,  
where $p$ is the ion momentum, $Q$ the charge and the $B$ the average 
field strength in the region where the ion propagates. For instance,
the effect of the ion bending for protons would cease at  $2.7 \times 10^{17}
eV$, while for iron this energy wlould be $1.5 \times 10^{19} eV$ in an average magnetic 
field of 3 $\mu$G
if $R$ is 100 $pc$, the nominal thickness of the gaseous disc.

These numerical games, though recurrent in the 
literature and textbooks, are both vague and thoroughly 
elusive to determine the exact positions of the $ankles$ along the
energy axis. In figure 10, the onset of the rectilinear propagation
for helium is shown as grammage versus energy. The rectilinear
propagation takes place at $4 \times 10^{18}$ $eV$.
The lack of a unique $R$ for all ions impedes the exact determination of the knee 
energies with the above equation. To each ion 
corresponds a $galactic$ $basin$ around the Earth with an intrinsic dimension [10,11]
dependent on the ion, and the correct estimate of the energy threshold
of the rectilinear propagation requires
the position of the solar cavity, the disc size and the nuclear cross sections.  

The mechanisms generating the ankles are the same 
as for the knees, but the physical conditions at the ankle energy region are less 
intricate. 

\section{THE NUCLEAR CROSS SECTIONS, ION ANKLES AND THE ANKLE}

Dissecting the phenomena which concur in the formation 
of the  $ankle$ and individual $ankles$ we are finally left with  
the role of the nuclear cross sections. 
The results reported 
in figures 4,6 and 7, along with those shown in the figure 11,  
suffice to highlight this role.  It is obvious from
the spectra shown in figure 7 how the function $n_g$, 
appearing in figure 4,  contributes to the formation of 
the ankle  (the spectral indices are incorporated in figure 7).
Imagine an ideal, unreal situation where the gap between the low and high plateau
is close to zero. Neither 
a change of slope in the spectrum above $ 6.0 \times 10^{15}$ $eV$ nor the ankle would have been observed. 
In this circumstance, the universal 
spectrum would have been entirely bound and regulated by the spectral indices, unaffected
by this unreal compression of $n_g$, with no $ankles$ at all. On
the contrary, with a  finite gap
between the ,low and high plateau,  the appearence of the ankle is a necessity.

The position and magnitude of the ankle and its necessity may be further
appreciated by taking into account:  
(1) the difference between the high and low plateau for the same ion; 
(2) the difference
between the low and high plateau in different ions; (3)  
the difference between low and high plateau for extragalactic cosmic rays.

The high plateau of any ion is fixed by the average matter thickness ($grammage$)
encountered by galactic cosmic rays and by $\sigma$. In the energy
region of the low plateau the physical
conditions present in the high plateau region are unchanged, except
for lower grammages (see figure 10) which ultimately imply low $n_g$ values. 
To elucidate the point (2) let us consider two different ions: He and 
Fe, for example. Since $\sigma$(Fe) is higher than $\sigma$(He),  galactic 
iron suffers severe losses in the disc with respect to He losses. Therefore
 $n_g$ for iron  becomes smaller than that 
for helium, a result displayed in figure 6; similarly for other ion pairs
with different $\sigma$. The low plateau
of iron is similar to that of helium, and all low plateaux
are almost equal for all ions (see figure 6). This is simply due to the 
exponential function 
governing nuclear collisions via the attenuation length $\lambda$ = 
$A$/$\sigma$$d$$N_A$. Since $d$ is extremely small, less than 0.008 $g$/$cm^2$,
the difference in the product $\sigma$$d$ for different ions is 
quite modest and the quantities $n_g$ for different 
ions cluster around a minimum value (see figure 6). By this argument it 
follows that the proton gap  must be higher than that of helium and the 
helium gap higher than that of iron etc.

A quite different result pertains to the
extragalactic component for which the grammage is higher than that of
the galactic component 
and the quantities $n_g$ 
in the energy band of the low plateau largely differ.
Above the energy of the rectilinear
propagation the relative intensities of all extradisc ions have to be in the 
same ratios as do
nuclear cross sections: $\sigma$(p)/$\sigma$(He),
 $\sigma$(Fe)/$\sigma$(He) etc.  This aspect is beautifully displayed in figure 11
above $10^{18}$ $eV$ for He, Fe and protons using the number of nuclear
collisions in the disc instead of the companion variable $n_g$.
 
The origin of the ankle described here  
may be further appreciated in comparison with other recent 
studies [26] where the ankle is a mere 
effect caused by the extragalactic component, introduced $ad$ $hoc$.

\end{document}